\documentclass[sigconf]{acmart}
\copyrightyear{2026}
\acmYear{2026}
\setcopyright{cc}
\setcctype{by}
\acmDOI{XXXXXXX.XXXXXXX}
\acmConference[CHASE 2026]{Proceedings of the 19th IEEE/ACM International Conference on Cooperative and Human Aspects of Software Engineering}{April 13--14, 2026}{Rio de Janeiro, Brazil}
\acmBooktitle{Proceedings of the 19th IEEE/ACM International Conference on Cooperative and Human Aspects of Software Engineering, April 13--14, 2026, Rio de Janeiro, Brazil}
\acmDOI{10.1145/XXXXXXX.XXXXXXX}
\acmISBN{979-8-4007-XXXX-X/2026/04}

\usepackage[utf8]{inputenc}
\usepackage{amsmath}

\usepackage{amssymb}
\usepackage{amsfonts}

\usepackage{array}
\usepackage{wrapfig}
\usepackage{multirow}
\usepackage{tabularx}
\usepackage{caption}
\usepackage{subcaption}
\usepackage{booktabs}
\usepackage{setspace}
\usepackage{tikz}

\usepackage{booktabs}
\usepackage{enumitem}
\usepackage{multirow}
\usepackage{longtable}
\usepackage{changepage}
\usepackage{hyperref}
\hypersetup{
    colorlinks=true,
    linkcolor=black,
    filecolor=black,      
    urlcolor=black,
    citecolor=black,
}

\usepackage{algorithmic}
\usepackage{graphicx}
\usepackage{textcomp}
\usepackage{xcolor}
\usepackage{booktabs}
\usepackage{booktabs}
\usepackage{enumitem}
\usepackage{tcolorbox}
\usepackage{enumitem}
\usepackage{fancybox}
\DeclareUnicodeCharacter{2212}{-}

\definecolor{background-gray}{gray}{0.96}

\usepackage[framemethod=tikz]{mdframed}
\newmdenv[
  topline=false,
  bottomline=false,
  rightline=false,
  skipabove=\baselineskip,
  skipbelow=\baselineskip,
  innertopmargin=6pt,
  innerbottommargin=6pt,
  innerleftmargin=12pt,
  innerrightmargin=12pt,
  tikzsetting={draw=black, line width=4pt},
  linecolor=black,
  backgroundcolor=background-gray
]{results}

\usepackage{fontawesome} 

\begin{document}

\title{How Does Cognitive Capability and Personality Influence Problem Solving in Coding Interview Puzzles?}

\author{Dulaji Hidellaarachchi}
%\authornote{}
\email{dulaji.hidellaarachchi@rmit.edu.au}
\orcid{0000-0003-0217-5317}
%\author{}
%\email{dulaji.hidellaarachchi@rmit.edu.au}
\affiliation{%
  \institution{RMIT University}
  \city{Melbourne}
  \state{Victoria}
  \country{Australia}
}

\author{Sebastian Baltes}
\email{sebastian.baltes@uni-heidelberg.de}
\affiliation{%
  \institution{Heidelberg University}
%  \city{}
  \country{Germany}}

\author{John Grundy}
\email{john.grundy@monash.edu}
\affiliation{%
 \institution{Monash University}
 %\streetaddress{Rono-Hills}
 \city{Melbourne}
 \state{Victoria}
 \country{Australia}}
%\author{}
%\affiliation{%
 % \institution{}
 % \city{}
 % \country{}
%}


\begin{abstract}
Software engineering involves cognitively demanding activities impacted by individual differences. We investigate how cognitive capability and personality traits are associated with software problem solving accuracy.
We assessed cognitive capability using Baddeley’s three-minute grammatical reasoning test. Personality was measured using the IPIP-NEO-50 test. Eighty participants (40 software practitioners and 40 software engineering students) completed these two tests with nine interview-style problem solving tasks, comprising six coding-related and three logical-reasoning questions. Our practitioners achieved slightly higher grammatical reasoning accuracy than students, although this difference was not statistically significant. Students achieved higher accuracy on the coding and logical-reasoning tasks. For all, grammatical reasoning accuracy was positively correlated with problem solving accuracy, indicating that individuals with higher reasoning accuracy tended to perform better on applied problem solving tasks. Conscientiousness was the strongest personality-related association, positively correlated with both grammatical reasoning and problem solving accuracy. Openness to experience was also positively correlated with grammatical reasoning and problem solving accuracy. Neuroticism showed a small negative correlation with problem solving accuracy and weak negative correlation with grammatical reasoning accuracy. 
Practical implications for education and industry include integrating structured reasoning tasks in curricula, and considering the interplay of personality and cognition in recruitment and role allocation.
%We highlight directions for future research, such as longitudinal and more task-diverse replications with larger samples.
\end{abstract}




\maketitle

\section{Introduction} \label{intro}
Software engineering (SE) is a cognitively demanding human-centered discipline in which software practitioners are heavily involved in problem solving and decision making at all stages of the software development life cycle \cite{curtis1988field,lenberg2015human, capretz2018call}. Activities such as eliciting and analyzing requirements, designing software, implementing new features, debugging, and communicating design decisions involve complex mental processes, including logical reasoning, abstraction, and attention management \cite{letovsky1987cognitive, siegmund2014understanding, mohanani2018cognitive, gonccales177towards}. 
This indicates that SE is not purely technical, but involves cognitively demanding activities that are shaped by how individuals perceive, interpret, and manage information under varying task constraints~\cite{fagerholm2022cognition,robillard1998measuring}. During the past decade, Human-Centric Software Engineering (HCSE) has emerged as a key research paradigm, recognizing the developer as a complex socio-technical actor rather than a purely technical resource \cite{pirzadeh2010human, hoda2021socio}. Based on this paradigm, researchers have examined how individual differences in motivation, personality, emotions, empathy, and cognition influence productivity, performance, creativity, and collaboration \cite{graziotin2014happy, cruz2015forty, kosti2014personality, fagerholm2022cognition}. For example, personality traits such as conscientiousness and openness to experience have been linked to persistence, adaptability, and creative thinking, while neuroticism has been associated with stress reactivity and fluctuating performance \cite{capretz2018call, cruz2015forty, mccrae1992introduction}. Despite this progress, the interaction between cognitive capability and personality remains underexplored in SE. Most of these existing studies treat these constructs separately rather than examining how they jointly influence SE.

Evidence from cognitive psychology and empirical SE suggests that reasoning ability and working memory  capacity support problem solving in programming/debugging tasks \cite{busjahn2015eye, arunachalam1996cognitive, bednarik2008studying}. At the same time, studies in SE show that non-cognitive attributes such as motivation and personality shape how practitioners approach and sustain effort in complex problem contexts \cite{graziotin2014happy}. Together, this indicates that cognitive capability and personality may operate jointly such as cognition enabling technical reasoning and personality supporting behavioral consistency, yet empirical studies investigating these relationships and quantifying both dimensions within the context of SE remain minimal. 
However, few studies have directly compared practitioners and students using standardized cognitive and personality measures. Previous work suggests that students often rely on structured analytical reasoning, whereas practitioners draw on experimental heuristics and contextual judgment \cite{begel2008novice, ko2007information}. Recent work also highlights the ongoing debate about the external validity of student-only studies and motivates direct comparisons between students and professionals where feasible \cite{falessi2018empirical}. Better understanding how these groups differ or align in cognitive capability and personality can inform both SE education and industry recruitment, clarifying whether academic cognitive indicators translate into professional performance.

From a conceptual perspective, practitioners and students may be expected to diverge due to differences in professional exposure and task familiarity. Practitioners are more likely to have encountered similar problem-solving tasks during technical interviews and daily work, while student performance may reflect more recent academic training with less exposure to applied reasoning contexts under time pressure~\cite{chedid2022university}.
Our study addresses these gaps by investigating the combined role of cognitive capability and personality in software problem-solving performance by collecting data from 40 software practitioners and 40 SE students. Participants completed the \emph{IPIP-NEO-50 personality test}~\cite{johnson2014measuring} and \emph{Baddeley Grammatical Reasoning Test}~\cite{baddeley19683}, as well as the same set of short coding and logical-reasoning questions modeled on coding interview tasks. We use interview-style problem solving accuracy as the comparative lens because coding and logic puzzles that require reasoning about algorithms, data structures, and pseudocode tracing are widely used to evaluate candidates in SE technical interviews. This provides a shared evaluative context that is familiar to practitioners and is increasingly encountered by students preparing for professional roles~\cite{mongan2012programming, aziz2019elements, laakmann2009cracking}.
This design allows us to compare cognitive capability and personality predictors between the practitioner and student groups under consistent task conditions. Accordingly, we aim to answer the following overarching research question:

\begin{quote}
\textbf{\emph{How do cognitive capability and personality traits together influence software problem-solving accuracy across practitioners and students?}}
\end{quote}

The collected data was analyzed using Microsoft Excel and Python, incorporating descriptive summaries, comparative statistics, and correlation analysis, and effect sizes. This enabled a detailed examination of how reasoning ability, personality traits, and problem-solving outcomes relate within and across participant groups. Our results show that students achieved higher accuracy in coding and logical-reasoning tasks. Higher reasoning ability was positively associated with problem-solving accuracy, while conscientiousness and openness to Experience emerged as positive correlates. In contrast, higher neuroticism was associated with slightly lower accuracy, suggesting that emotional regulation may be associated with cognitive efficiency. Our research makes the following key contributions:
\begin{itemize}
    \item a reusable empirical study setup for jointly examining cognitive and personality characteristics across students and practitioners using validated instruments;
    \item novel insights into how cognitive capability and personality jointly influence software problem-solving accuracy; and
    \item implications for software education, recruitment, and training, emphasizing that both cognitive and personality indicators can complement technical assessments in evaluating software development potential.
\end{itemize}

\section{Motivation} \label{background}

Research increasingly recognizes that software development is not purely technical, but is shaped by how individuals think, reason, and act within complex environments. Software engineers engage in tasks that require continuous interpretation and decision-making such as debugging, requirements analysis, and design reasoning, which are highly dependent on cognitive processing and individual differences~\cite{robillard1998measuring, hidellaarachchi2021effects, detienne2001software}. Understanding these human aspects provides valuable insights into how people perform in software projects and how education can better prepare future practitioners.

Personality influences how individuals collaborate, communicate, and approach technical work~\cite{navarro2024individual}. Studies adopting the \emph{Five-Factor Model (FFM)} have linked personality traits to diverse set of characteristics. For example, conscientiousness refers to characteristics such as the hard-working, organized, reliable, responsible nature of individuals, while higher neuroticism has been associated with lower satisfaction and teamwork challenges~\cite{capretz2003personality, cruz2015forty, feldt2010links, salleh2009empirical}. Much empirical work has explored personality across different roles from developers and testers to analysts, highlighting its relevance to productivity, motivation, and defect detection~\cite{kanij2015empirical, hannay2009effects, john1999big, hidellaarachchi2024impact}. However, most of these studies have examined personality through self-report questionnaires, without connecting it to measurable reasoning or problem-solving outcomes. This leaves a gap in understanding how these characteristics manifest in actual software-related thinking processes. 

SE tasks involve sustained reasoning and abstraction~\cite{siegmund2014measuring, pennington1987stimulus}. Controlled experiments show that contextual factors such as environmental noise, visual representation, or model format can directly affect requirement comprehension and analytical accuracy~\cite{romano2018effect, turner2014eye, ricca2014assessing}. For example, adding diagrams or mock-ups to textual specifications improves understanding, whereas noisy conditions reduce comprehension speed and precision~\cite{romano2018effect, ricca2014assessing}. Eye-tracking and comprehension studies further demonstrate that experienced engineers adopt more efficient reasoning patterns than novices, reflecting differences in cognitive control and flexibility~\cite{abbad2022estimating,sharafi2015systematic,turner2014eye}. Cognitive capability therefore represents a critical element in the way engineers handle ambiguity, integrate information, and make decisions under time or complexity constraints~\cite{baddeley19683, almulla2023integrated}. 

Problem solving is central to SE, from interpreting requirements to design, implementation, and testing. These tasks require both logical and conceptual reasoning, with successful solutions depending on how individuals structure and evaluate information~\cite{lister2004multi}. Novices often rely on surface features of problems, whereas professionals use experience-driven heuristics and pattern recognition to navigate complexity~\cite{siegmund2014measuring}. Studies in software comprehension and testing show that such reasoning differences are closely tied to performance outcomes, including task accuracy, completion time, and error detection~\cite{kanij2015empirical, kanij2013investigation, romano2018effect}. However, few studies have systematically examined how personality traits and cognitive capability jointly influence this performance.

Building on the existing body of research, our study examines the relationship between personality, cognitive capability, and problem-solving accuracy among practitioners and students. By combining validated psychological instruments such as a standard personality test with reasoning and problem-solving tasks, our study moves beyond isolated measures of human traits toward a holistic understanding of how individual differences affect analytical performance in software contexts. Comparing students and practitioners provides insight into how personality and cognitive capability interact in shaping problem-solving accuracy. The results contribute to ongoing HCSE discussions on integrating human aspects into SE research and education, highlighting that improving software practice also requires understanding the people who perform it.

\section{Research Methodology}
We employed a quantitative, cross-sectional survey design to examine relationships between cognitive capability, personality, and problem-solving accuracy among practitioners and students. A survey-based approach enabled standardize and consistent collection of individual measures, background information, and task responses within a single instrument \cite{creswell2017research, fowler2013survey}. %This allows a consistent administration and supports scalable data collection while maintaining standardization of instructions, ordering, and task conditions between practitioners and students.
Our study was approved by Monash University Ethics Committee; approval ID:  40008.

\subsection{Survey Design and Pilot Study} \label{survey design}
\subsubsection{Survey Structure}
Our survey comprised four sections; \textit{Demographics} to capture details such as participants' role, years of experience and educational background, \textit{Personality Test} to measure the \emph{Big Five} personality traits using the \emph{IPIP-NEO 50} scale, \emph{Baddeley's Grammatical Reasoning Test (GRT)} to assess cognitive capabilities and \emph{coding and logical-reasoning questions} to use in evaluating applied reasoning in an SE context. The survey was implemented and administered via Qualtrics, with all questions presented in a fixed sequence. Built-in timing controls ensured the grammatical reasoning section was limited to three minutes. 

\subsubsection{Pilot Study}
Three pilot surveys were conducted to assess the clarity of the questions, the flow of the survey, and the timing. Two pilots were conducted with practitioners currently employed in the IT industry and one with an academic who had previous professional software development experience. Feedback from these pilots led to refinements such as adjusting the coding and logical-reasoning questions, simplifying the instructions, and optimizing the order of the survey sections for smoother completion. 

\subsubsection{Personality Test}
Personality traits were measured using the \emph{International Personality Item Pool (IPIP) NEO-50} scale developed by Johnson, based on the Five-Factor Model of personality~\cite{johnson2014measuring}. This widely recognized test is based on the \emph{Five-Factor Model} (FFM) of personality, a well-established framework in psychology and frequently used in SE studies~\cite{kanij2013investigation, hidellaarachchi2024s}. The IPIP-NEO-50 includes 50 items, 10 per trait. The five main traits are~\cite{RN1619, hidellaarachchi2024s}:

\par \faSearchPlus\hspace{0.1cm}{\textbf{Openness to Experience:} individuals' intellectual, cultural or creative interests.  High-scored individuals tend to be imaginative, broad-minded and curious. Individuals scoring low tend to prefer familiarity and routine, place greater value on conventional approaches, and show less interest in abstract or novel experiences.}
\par \faSearchPlus\hspace{0.1cm}{\textbf{Conscientiousness:} refers to individuals' focus on achievements. High-scored individuals tend to be hardworking, organized, able to complete tasks thoroughly on time, and reliable. Low-scored individuals tend to be irresponsible, impulsive and disorganized.}
\par \faSearchPlus\hspace{0.1cm}{\textbf{Extraversion:} relates to the degree of sociability, activeness, talkativeness, and assertiveness.  The opposite end of this spectrum shows a lack of social involvement, shyness, and prefers to be alone more than extraverted people. This does not mean that they are unfriendly or antisocial; rather, they are reserved in social situations}.
\par \faSearchPlus\hspace{0.1cm}{\textbf{Agreeableness:} refers to positive traits such as cooperativeness, kindness, trust and warmth. Low-scored individuals on agreeableness tend to be skeptical, selfish and hostile.}
\par \faSearchPlus\hspace{0.1cm}{\textbf{Neuroticism:} refers to the state of emotional stability of individuals. Low-scored individuals tend to be calm, confident and secure, whereas high-scored individuals on neuroticism tend to be moody, anxious, nervous and insecure.}

Participants indicated how accurately they thought each statement described them on a five-point Likert scale ranging from `very inaccurate' (1) to `very accurate' (5). For example, a statement such as \textit{``I have a vivid imagination"} corresponds to the \textbf{Openness to Experience} trait, while \textit{``I am always prepared"} relates to \textbf{Conscientiousness} trait. Each item contributes to one of the five personality traits, providing a numerical profile that reflects the dominant personality tendencies of the participant. \emph{IPIP-NEO-50} was chosen for this study because it is openly available\footnote{\url{https://ipip.ori.org/New_IPIP-50-item-scale.htm}} and easy to administer and has strong psychometric reliability.
%This personality test was delivered online as part of the main survey, alongside demographic questions to contextualize participants’ profiles. % A copy of the full IPIP-NEO-50 item set used in this study is available in the Appendix A and B. -- repeats below

\subsubsection{Baddeley’s Grammatical Reasoning Test (GRT)} \label{grammatical_test}
Cognitive reasoning ability was assessed using \emph{Baddeley's Grammatical Reasoning Test} (GRT)~\cite{baddeley19683}, a well-established psychometric measure originally developed as a brief verbal reasoning task. The test presents 64 short statements that describe spatial relationships between two letters (e.g., ``A follows B''), requiring participants to decide whether each statement is true or false. The participants had three minutes to complete as many statements as possible, the number of correct responses being used as their cognitive reasoning score. We computed grammatical reasoning accuracy as (\% correct / attempted) and used this as our primary measure of cognitive capability. 
This test has been shown to reliably assess analytical reasoning and working memory under time pressure and continues to be used in modern adaptations such as the \emph{Mini-Q Intelligence Screening Test}~\cite{chamorro2008personality, baudson2016mini}, which demonstrates strong reliability and validity as a quick measure of speeded reasoning. The GRT is particularly suited for contexts where cognitive ability serves as a control or predictor variable, such as software problem solving, because it captures reasoning accuracy in a short time window without relying on domain-specific knowledge. More broadly, general cognitive reasoning ability has been shown to influence a wide range of downstream analytical activities, including comprehension, problem solving, and evaluation tasks, many of which are central to software development practice. As such, using the GRT enables an examination of how general reasoning ability relates to software problem-solving accuracy without conflating cognitive ability with prior technical knowledge or domain experience~\cite{begel2008novice}.

\subsubsection{Coding and Logical-Reasoning Questions} \label{coding_logical_questions}

Applied reasoning and technical problem-solving ability were assessed using nine short questions: six coding-related and three logical-reasoning items. The same questions were administered unchanged to all participants, with each question having a single correct answer. Overall problem-solving accuracy was computed as the total number of correct responses (range 0–9).
The questions were inspired by commonly used software engineering interview-style tasks and adapted from freely available, industry-oriented resources (e.g.,
\emph{GeeksforGeeks}\footnote{\url{https://www.geeksforgeeks.org/}}), 
which are widely used by both students and practitioners for interview preparation and skills assessment. Rather than assessing proficiency in a specific programming language or syntax, the questions were designed to capture general reasoning about program behavior, data structures, and algorithmic outcomes typically encountered in early-stage technical screening and problem-solving contexts \cite{aziz2019elements, laakmann2009cracking, mongan2012programming}.

The six coding-related questions assessed reasoning about program behavior and basic data-structure concepts. These included code comprehension and pseudocode-tracing tasks that involved interpreting program logic and following control flow through conditional statements and loops. In addition, some questions examined reasoning about fundamental data structures, such as understanding the effects of sequences of operations on stacks/queues, recognizing full and empty conditions in a circular queue, and tracking the state of a hash table after a series of insertions. Together, these questions required step-by-step reasoning about how algorithms and data structures behave under clearly specified rules. 
The three logical-reasoning questions were included to capture abstract reasoning demands that commonly appear in technical screening contexts but are not tied to programming knowledge. These involved number-series reasoning and short logic-based problems that required participants to infer patterns, apply constraints, and reason about numerical or relational relationships. Question selection was guided by discussions among the authors to ensure coverage of multiple reasoning demands (e.g., code comprehension, data-structure reasoning, and abstract logical reasoning), rather than to optimize task difficulty or maximize performance differences between software practitioners and students. The complete set of survey materials, including the personality test, GRT and coding and logical reasoning questions are provided in our supplementary materials \cite{hidellaarachchi_2026_18343984, hidellaarachchi_2026_18344007}. %A\footnote{\url{https://zenodo.org/records/18343984}~(Appendix A)} and B\footnote{\url{https://zenodo.org/records/18344007}~(Appendix B)}.

\subsection{Data Collection}
The study was conceptualized in late 2023 and implemented in two phases to compare software practitioners and students. Initial recruitment was conducted through the authors' professional network and social media platforms (e.g., LinkedIn and X/Twitter). Due to low response rates, recruitment was later expanded through the \emph{Prolific} research platform to reach verified industry professionals working in SE-related roles. We used filtering options in Prolific to reach out to our target participants. Practitioner data was collected between January and June 2024. Following the practitioner phase, a comparable survey was conducted with SE students, mirroring the same structure, instructions, and timing. Student data collection then took place in early 2025 via Prolific. All surveys were administered through Qualtrics, which controlled timing and automatically recorded responses. 

The survey took approximately 20-30 minutes to complete. Participants were clearly informed about the nature of the study and implied consent was obtained before starting the survey. In total, 80 participants (40 software practitioners, 40 students) completed the survey. All participants completed the survey in a single online session. Navigation across survey sections was permitted except for the GRT, which was strictly time-limited to three minutes and enforced by the survey platform. Once the time limit elapsed, the participants automatically advanced and could not return to the GRT. Following the GRT, participants proceeded to the coding and logical-reasoning questions, which were not time-restricted. This survey flow and timing configuration were identical for both practitioners and students.

\subsection{Data Analysis}
All data collected was analyzed using Excel and Python (utilizing the \texttt{pandas}, \texttt{numpy}, and \texttt{matplotlib} libraries), aiming to examine relationships between personality traits, cognitive capability, and problem-solving accuracy, as well as to compare these measures between practitioners and students. Data were first screened for completeness and internal consistency across the four survey sections. Only participants who completed all sections were included in the analysis (n = 80). For each participant, the dataset comprised demographic information, scores for the five personality traits, grammatical reasoning accuracy, and problem-solving accuracy derived from the coding and logical-reasoning questions.

Personality trait scores were computed following the standard IPIP-NEO-50 scoring guidelines, summing responses for positively and negatively keyed items associated with each trait. Cognitive capability was operationalized as grammatical reasoning accuracy, calculated as the percentage of correct responses out of the number of attempted statements within the three-minute time limit. Problem-solving accuracy was calculated as the total number of correct responses across the nine coding and logical-reasoning questions. Descriptive analyses were conducted separately for practitioners and students. Summary statistics including mean, median, standard deviation, and observed minimum and maximum values were computed for each personality trait, grammatical reasoning accuracy, and problem-solving accuracy. These descriptive measures provided an overview of central tendencies and variability within each group.

To compare practitioners and students, inferential comparisons were performed using Welch’s independent -samples $t$-tests. It was selected due to its robustness to unequal variances and potential distributional differences between groups. Alongside p-values, Cliff’s delta ($\Delta$) was reported as a non-parametric effect size measure to quantify the magnitude and direction of group differences in personality traits, grammatical reasoning accuracy, and problem-solving accuracy. This combination allows for interpretation of both statistical significance and practical relevance.

Finally, Pearson correlation analyses were conducted on the combined sample to examine associations between personality traits, grammatical reasoning accuracy, and total problem-solving accuracy at the individual level. These analyses were used to explore whether individual differences in personality and cognitive capability were associated with variation in problem-solving accuracy. Correlation coefficients and their statistical significance were reported, consistent with the exploratory aims of the study.
%The combined use of Excel and Python allowed for transparent data inspection and reproducible computation. All analytical steps were systematically documented, supporting reliability and traceability of the findings, which were interpreted in relation to existing literature on cognitive and personality influences in SE.
The anonymized dataset is provided in Appendix C \cite{hidellaarachchi_2025_17429307}
%\footnote{\url{https://doi.org/10.5281/zenodo.17429307}~(Appendix C)}.

\section{Findings}

\subsection{Participant Demographics}
Table \ref{TABLE 1: Participants' demographics-Practitioners} and Table \ref{TABLE 2: Participants' demographics-Students} summarize the demographic profiles of the practitioners (n = 40) and students (n = 40) who participated in the study. Both cohorts were comparable in size, but differed in age, educational background, and professional experience.

\begin{table}[t]
\centering
\caption{Participant demographics: software practitioners. %{\color{red} Please double-check and update the values; check that the examples in the parenthesis list the most common answers}
}
\label{TABLE 1: Participants' demographics-Practitioners}

\setlength{\tabcolsep}{4pt}
\footnotesize

\begin{tabularx}{\columnwidth}{@{}X r@{}}
\toprule

\multicolumn{2}{@{}p{\columnwidth}@{}}{\textbf{Education}} \\
\midrule
Bachelor’s degree & 62.5\% \\
Postgraduate degree (Master’s or Doctoral) & 30\% \\
Below bachelor’s level & 7.5\% \\

\midrule
\multicolumn{2}{@{}p{\columnwidth}@{}}{\textbf{Job Roles}} \\
\midrule
Software engineers/developers (incl. junior/senior/specialist) & 57.5\% \\
Management roles (project/IT managers) & 17.5\% \\
Testing and QA roles & 10\% \\
Other IT roles & 15\% \\

\midrule
\multicolumn{2}{@{}p{\columnwidth}@{}}{\textbf{Country of Residence}} \\
\midrule
United States & 25\% \\
Europe (Poland, France, UK, others) & 47.5\% \\
Africa (South Africa) & 12.5\% \\
Other regions (Canada, Australia, others) & 15\% \\

\midrule
\multicolumn{2}{@{}p{\columnwidth}@{}}{\textbf{Primary Work Domain}} \\
\midrule
Technology & 77.5\% \\
Non-technology sectors (education, finance, health, others) & 22.5\% \\

\midrule
\multicolumn{2}{@{}p{\columnwidth}@{}}{\textbf{Programming Languages}} \\
\midrule
Web languages (JavaScript, HTML/CSS, TypeScript) & 15--62.5\% \\
General-purpose languages (Python, Java, C\#, C++) & 20--37.5\% \\
Other languages/tools (e.g, SQL, Bash, R) & 5--47.5\% \\

\bottomrule
\end{tabularx}
\end{table}

\begin{table}[t]
\centering
\caption{Participant demographics: students.}
\label{TABLE 2: Participants' demographics-Students}

\footnotesize

\begin{tabularx}{\columnwidth}{@{}X r@{}}
\toprule

\multicolumn{2}{@{}p{\columnwidth}@{}}{\textbf{Current Degree Program}} \\
\midrule
Bachelor’s programs in computing (CS, SE, IT, double degrees) & 95\% \\
Other bachelor’s programs & 5\% \\

\midrule
\multicolumn{2}{@{}p{\columnwidth}@{}}{\textbf{Country of Residence}} \\
\midrule
United States & 22.5\% \\
Europe (UK, Germany, Portugal, Poland, Denmark, Italy, others) & 62.5\% \\
Other regions (Canada, New Zealand, Mexico, Chile, others) & 15\% \\

\midrule
\multicolumn{2}{@{}p{\columnwidth}@{}}{\textbf{Programming Languages}} \\
\midrule
Web languages (JavaScript, HTML/CSS, TypeScript) & 27.5-80\% \\
General-purpose languages (Python, Java, C\#, C++) & 30--72.5\% \\
Other languages/tools (e.g, SQL, Bash, R) & 10--75\% \\

\bottomrule
\end{tabularx}
\end{table}

Software professionals represented a diverse group that spans multiple roles and industries. Most practitioners were between 25-44 years old (70\%), with a smaller proportion aged 18–24 (22.5\%) and only three participants older than 45. The majority identified as men (67.5\%), followed by women (27.5\%), with 5\% identifying as non-binary/ preferring not to disclose. More than 90\% of practitioners held at least a bachelor’s degree, including 30\% of them with a postgraduate degree (Master's or Doctoral). Most practitioners reported between one and nine years of experience in the SE industry (80\%), while a further 15\% reported more than ten years of experience, confirming that the practitioner sample consisted of experienced professionals. Practitioners occupied a range of professional roles, with the majority working in SE/development positions (57.5\%), alongside representation from project/IT management (17.5\%), testing and quality assurance (10\%), and other IT-related roles. Their professional work spanned multiple application domains, predominantly technology (77.5\%), with additional representation from non-technology sectors such as education, finance, and health. Practitioners reported experience with multiple programming languages, most commonly web technologies (e.g., JavaScript, HTML/CSS), general-purpose languages (e.g., Python, Java, C\#), and other tools such as SQL (Table \ref{TABLE 1: Participants' demographics-Practitioners}). The practitioner sample was geographically diverse, with participants from the United States, Europe, Africa, and other regions. 

Regarding linguistic background, 67.5\% of practitioners identified as native English speakers. Among non-native speakers, most reported advanced or native-like English proficiency (12 out of 13). Given the language-dependent nature of the grammatical reasoning task, variation in English proficiency, although limited, may have influenced performance. A similar consideration applies to the student group.
 %Given the language-dependent nature of the grammatical reasoning task, variation in English proficiency among non-native participants, although minimal, may have influenced performance. A similar condition applies to the student group as well.
The student cohort consisted primarily of undergraduate students enrolled in computing-related degree programs, predominantly computer science and SE (Table \ref{TABLE 2: Participants' demographics-Students}). The majority were between 18–24 years old (55\%), representing a typical demographic of universities, and the rest were in the age range of 25 to 34. The majority (75\% ) was men, followed by 25\% of women. Students were geographically diverse, with participants from Europe (62.5\%), United States (22.5\%) and other regions (e.g, Canada, New Zealand etc. (15\%)). Language proficiency was comparable to the practitioner group, with 65\% identifying as native English speakers and most non-native speakers reporting advanced fluency. Although classified as students, 42.5\% reported prior or current industry exposure through employment, internships, or previous roles, typically at an early-career level. Students also reported familiarity with a broad range of programming languages, including web technologies, general-purpose languages, and other tools, closely aligning with the categories reported by practitioners (Table \ref{TABLE 2: Participants' demographics-Students}).

\begin{table*}[t]
\centering
\footnotesize
\caption{Descriptive statistics and group comparisons of personality traits for practitioners and students.}
\label{personality_stats}
\begin{tabular}{p{3.2cm}lccccc c}
\toprule
\textbf{Personality Trait} &
\textbf{Group} &
\textbf{Mean} &
\textbf{Median} &
\textbf{SD} &
\textbf{Min--Max} &
\textbf{$p$} &
\textbf{Cliff’s $\Delta$} \\
\midrule
Extraversion
 & Practitioners & 27.73 & 28.00 & 9.35 & 10--46 & 0.463 & -0.09 \\
 & Students      & 29.03 & 29.00 & 6.07 & 14--42 &       &       \\
\midrule
Agreeableness
 & Practitioners & 36.50 & 37.00 & 6.47 & 22--49 & 0.260 & -0.14 \\
 & Students      & 37.95 & 38.00 & 4.82 & 25--47 &       &       \\
\midrule
Conscientiousness
 & Practitioners & 36.63 & 37.00 & 7.26 & 20--49 & 0.307 & +0.15 \\
 & Students      & 35.03 & 35.00 & 6.64 & 21--48 &       &       \\
\midrule
Neuroticism
 & Practitioners & 32.65 & 33.00 & 7.49 & 18--49 & 0.701 & +0.07 \\
 & Students      & 31.95 & 32.00 & 8.69 & 14--48 &       &       \\
\midrule
Openness to Experience
 & Practitioners & 38.48 & 39.00 & 5.98 & 23--50 & 0.617 & +0.05 \\
 & Students      & 37.80 & 38.00 & 6.04 & 22--49 &       &       \\
\midrule

\multicolumn{8}{p{\columnwidth}}{\footnotesize $p$-values from Welch’s $t$-tests; Cliff’s $\Delta$ reported as effect size (Practitioners -- Students).} \\
\bottomrule
\end{tabular}

\end{table*}

\subsection{Personality Profiles}
\begin{figure}[]
  \includegraphics[width=\columnwidth]{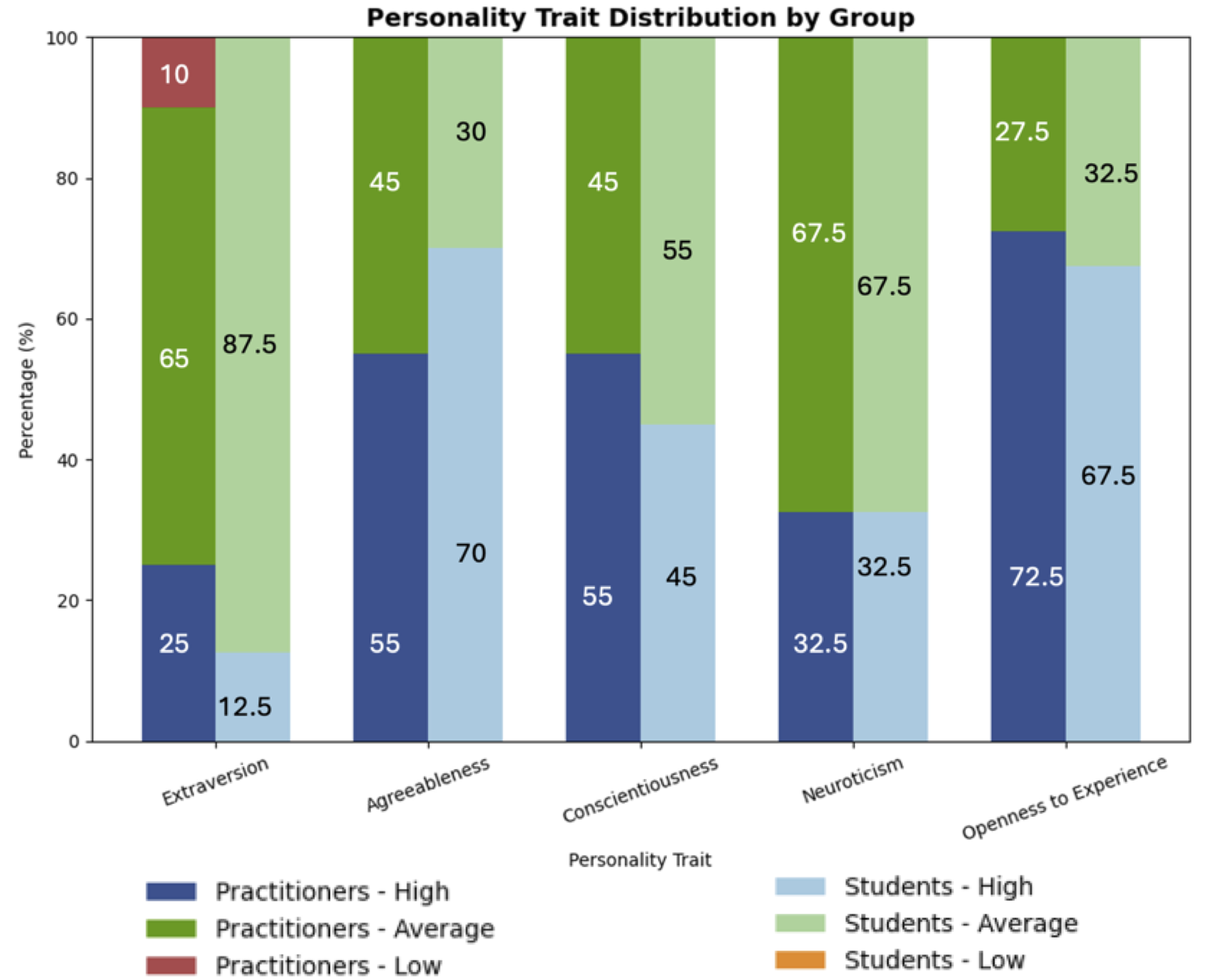}
 \caption{Personality traits among software practitioners and students.}
  \label{distribution}
\end{figure}

Figure \ref{distribution} summarizes the personality profiles of software practitioners and students. The percentage score of each participant obtained for each trait was classified as high (70\% or greater), average (31–69 \%), or low (30 \% or below), following the standard classification of the personality test. 
Among software practitioners (n = 40), openness to experience emerged as the most dominant trait. Nearly three-quarters (72.5 \%) of software practitioners scored within the \textit{high} range, indicating their imaginative, broad-minded and curious nature, the attributes which are closely aligned with software innovation and problem solving. Conscientiousness and agreeableness were the next most pronounced traits (55\% high each), \textcolor{black}{highlighting practitioners' tendency toward organization, hard-working nature, cooperative behaviour, which are commonly associated with task completion, process adherence and self-discipline in professional contexts \cite{barrick1991big}}. Referring to the extraversion trait, the majority (65\%) obtained average scores, where 25\% obtained high scores. Neuroticism scores were also mostly average (67.5\%), with 32.5\% high and no low scores. This moderate spread indicates variability in emotional stability among practitioners, without suggesting extreme levels at either end of the distribution. 
%This moderate spread could suggest that practitioners generally manage stress well but differ in emotional sensitivity depending on role and experience. 
Overall, practitioners displayed a personality composition characterized by openness to experience, conscientiousness and agreeableness as dominant, complemented by but moderately extroverted interpersonal behaviour.

%\begin{figure}[t]
 % \includegraphics[width=0.8\columnwidth]{Figure-Personality Traits - Participants.pdf}
  %\caption{Summary of personality profiles of the Practitioners}
 % \label{PersonalityProfilesPractitioners}
%\end{figure}

A broadly similar pattern was observed among students (n = 40). Agreeableness was the most dominant trait with 70\% of students scoring in the high range, indicating cooperative and supportive tendencies that may facilitate group learning and collaboration during academic projects. This was followed by openness to experience (67.5\%), reflecting tendencies toward curiosity, imagination and openness to new ideas. Conscientiousness was moderately present (45\% High), highlighting characteristics such as hard-working and organized nature, while extraversion was mostly average (87.5\%). Neuroticism among students followed a similar pattern to software practitioners, with most participants scoring in the average range (67.5\%) and 32.5\% high, suggesting that while many students are emotionally stable, a subset experiences greater stress or performance-related anxiety typical in academic settings.

%In addition to these categorical distributions of the personality profiles of both groups, 
Table \ref{personality_stats} summarizes the descriptive and inferential statistics for the five personality traits across both groups. It presents measures of central tendency and dispersion (mean, median, standard deviation (SD), and observed range), alongside between-group comparisons using Welch's \textit{t}-tests and corresponding effect sizes measured with Cliff’s $\Delta$. This indicates that software practitioners and students share similar mean rankings across traits, with openness to experience and agreeableness consistently scoring highest in both groups. However, software practitioners exhibit greater variability for several traits, particularly extraversion, conscientiousness and agreeableness as reflected in larger standard deviations and wider observed ranges compared to students. In contrast, variability in openness to experience is comparable across groups, while neuroticism shows slightly greater variability among students. Despite these differences in dispersion, Welch's \textit{t}-tests revealed no statistically significant differences between practitioners and students for any of the five personality traits. Correspondingly, Cliff’s $\Delta$ values were negligible to small across all traits, indicating substantial overlap between practitioner and student personality distributions. 

This participant personality analysis suggests that software practitioners and SE students share broadly similar personality profiles characterized by high openness to experience and conscientiousness-related tendencies, while differing primarily in the degree of variability rather than central tendency. Students exhibit more homogeneous profiles, particularly for agreeableness, whereas practitioners demonstrate greater dispersion across several traits, reflecting more differentiated personality patterns.\\

\begin{results}
\textbf{Personality:}
Practitioners and students displayed broadly similar personality profiles, with openness to experience and agreeableness showing the highest scores in both groups, followed by conscientiousness. While practitioners exhibited greater variability for several traits, particularly extraversion, conscientiousness, and agreeableness, students showed more homogeneous profiles, especially for agreeableness. Despite these differences in dispersion, no statistically significant differences were observed between practitioners and students across any of the five personality traits. 
\end{results}

%The overall personality across both groups aligns strongly with the profiles commonly associated with effective software professionals~\cite{hidellaarachchi2024s, hidellaarachchi2024impact}: high openness to experience and conscientiousness, average in extraversion, and predominantly average neuroticism (Figure~\ref{distribution}). Openness to experience stands out as the most defining trait in both groups, with means exceeding 78\%. 
%The slightly higher values of openness to experience (mean = 81.4\%) of practitioners may reflect the continued need to adapt to emerging technologies, frameworks, and problem solving contexts in professional practice.

%\begin{figure}[t]
%  \includegraphics[width=0.8\columnwidth]{Figure_Students_Personality.png}
 % \caption{Summary of personality profiles of the Students}
%  \label{PersonalityProfilesStudents}
%\end{figure}

\subsection{Cognitive Capabilities Assessment}

\begin{table}[b]
\centering
\footnotesize
\caption{Comparison of grammatical reasoning accuracy between practitioners and students.}
\label{comparison_table}
\begin{tabular}{lrr}
\toprule
\textbf{Measure} & \textbf{Practitioners} & \textbf{Students} \\
\midrule
Statements Attempted (Mean $\pm$ SD)      & 37.4 $\pm$ 17.2  & 37.5 $\pm$ 13.9 \\
Accuracy (Mean $\pm$ SD, \%)              & 76.1 $\pm$ 22.0  & 70.9 $\pm$ 15.9 \\
95\% Confidence Interval (Accuracy, \%)   & [69.1, 83.1]     & [65.8, 75.9] \\
Median Accuracy (\%)                      & 84.5             & 75.0 \\
Accuracy Range (\%)                       & 10.0--100.0      & 30.8--96.2 \\
\midrule
\multicolumn{3}{l}{\textbf{Welch’s $t$-test (accuracy): $p = 0.227$}} \\
\multicolumn{3}{l}{\textbf{Effect size: Cliff’s $\Delta = 0.27$ (small)}} \\

\bottomrule
\end{tabular}
\end{table}

The cognitive capabilities of the participants were measured using Baddeley's Grammatical Reasoning Test (GRT), which evaluates the precision of reasoning and the processing speed under time pressure (see Section \ref{grammatical_test}). 
%In the survey, each participant was presented with 64 short statements describing spatial relationships between letters (e.g., ``A follows B'') and asked to respond `true' or `false' within three minutes. 
Accuracy was calculated as the percentage of correct responses out of the number of statements attempted. For between-group comparisons on continuous accuracy measures, Cliff’s $\Delta$ was used as a distribution-free effect size to quantify the magnitude and direction of group differences.

\begin{figure}[]
  \includegraphics[width=0.95\columnwidth]{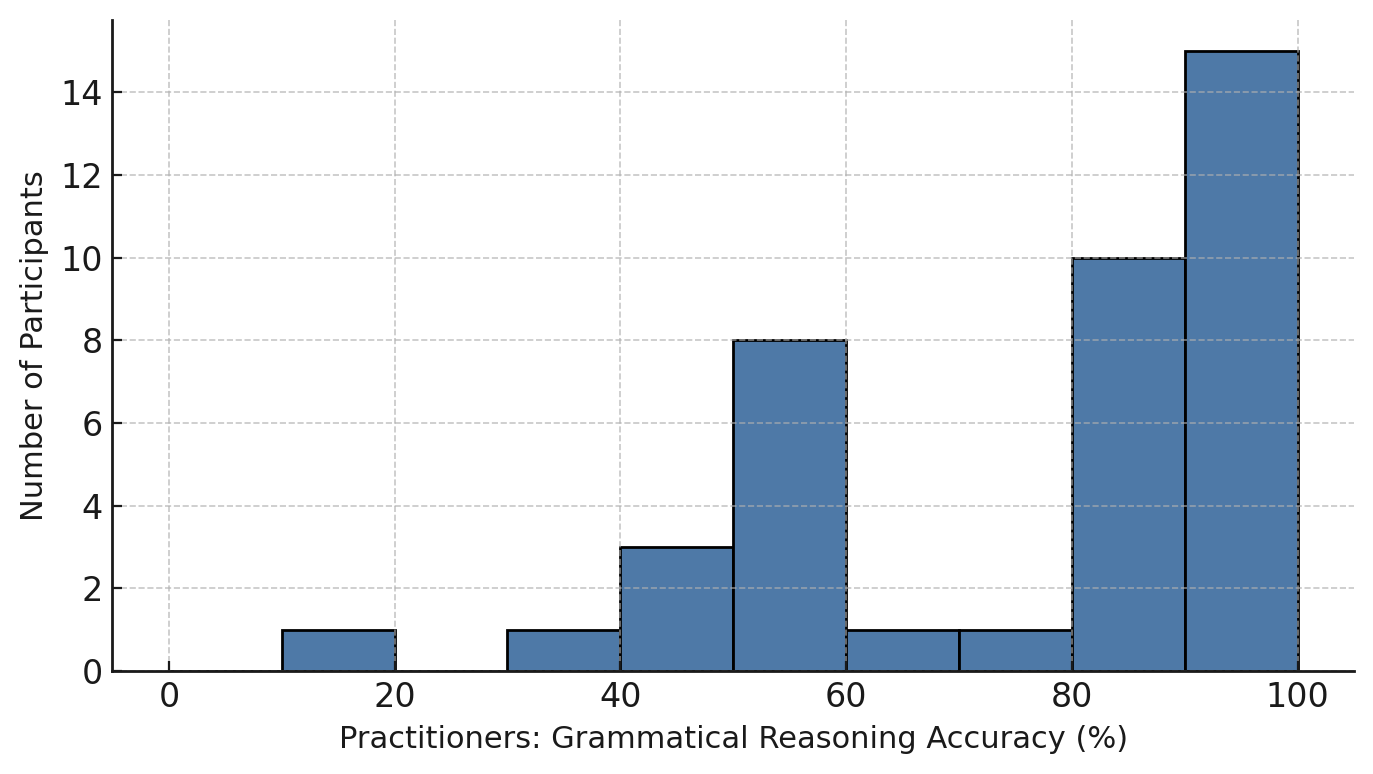}
  \caption{Grammatical reasoning accuracy of practitioners.}
  \label{His_Prac}
\end{figure}

\begin{figure}[]
  \includegraphics[width=0.95\columnwidth]{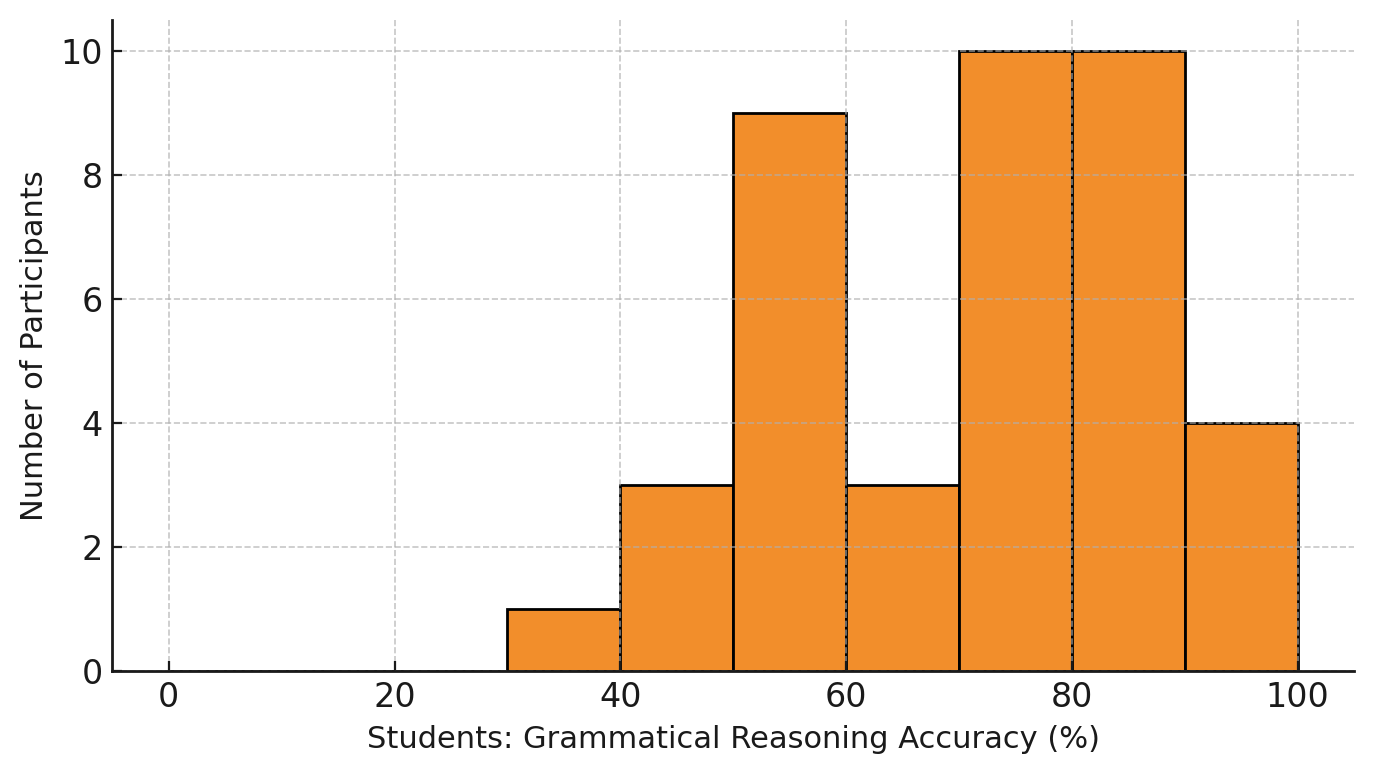}
  \caption{Grammatical reasoning accuracy of students.}
  \label{His_Stu}
\end{figure}

Descriptive results for grammatical reasoning accuracy are summarized in Table \ref{comparison_table}. Practitioners attempted an average of 37.4 statements (SD = 17.2) within the three-minute limit and achieved a mean accuracy of 76.1\% (SD = 22.0). Students attempted a comparable number of statements (M = 37.5, SD = 13.9) but achieved a slightly lower mean accuracy of 70.9\% (SD = 15.9). Median accuracy further reflects this difference, with practitioners achieving a median accuracy of 84.5\% compared to 75.0\% among students. Although both groups demonstrated relatively high grammatical reasoning accuracy overall, practitioners exhibited a wider observed accuracy range (10.0–100.0\%) compared to students (30.8–96.2\%). In particular, several practitioners achieved near-perfect accuracy, whereas student accuracy scores were more tightly clustered.

The 95\% confidence intervals provide additional context for central tendency and spread. Practitioners’ accuracy confidence interval ([69.1\%, 83.1\%]) sits slightly above that of the students ([65.8\%, 75.9\%]), indicating a modest difference in central tendency. However, Welch's independent-samples \textit{t}-test showed that this difference in grammatical reasoning accuracy was not statistically significant ($p = 0.227$). Consistent with this result, the effect size was small (Cliff’s $\Delta = 0.27$ ), indicating substantial overlap between the accuracy distributions of the two groups.

%\begin{table}[t]
%\centering
%\caption{Descriptive statistics of participants’ grammatical reasoning accuracy}
%\resizebox{\columnwidth}{!}{%
%\begin{tabular}{lccccccc}
%\hline
%\textbf{Cohort} & \textbf{n} & \textbf{Answered} & \textbf{Correct} & \textbf{Accuracy} & \textbf{Median} & \textbf{Min} & \textbf{Max} \\
% &  & \textbf{Mean (SD)} & \textbf{Mean (SD)} & \textbf{Mean (SD)} &  &  &  \\
%\hline
%Practitioners & 40 & 39.0 (15.6) & 29.8 (14.7) & \textbf{76.1 (12.5)} & 78.9 & 9.1 & 100 \\
%Students      & 40 & 38.2 (12.4) & 27.1 (11.6) & \textbf{71.0 (10.7)} & %73.9 & 30.8 & 95.8 \\
%\hline
%\end{tabular}%
%}
%\label{grammar_stats}
%\end{table}

The accuracy distributions for both cohorts are shown in Figure \ref{His_Prac} and Figure \ref{His_Stu}. Practitioners exhibit a broader and slightly right-skewed distribution, with most accuracy scores concentrated between 70\% and 90\% and several individuals achieving very high accuracy approaching 100\%. This extended upper tail highlights greater dispersion in grammatical reasoning accuracy within the practitioner group. In contrast, students display a more compact and approximately symmetric distribution centered around the 70–80\% range, with fewer observations at the extremes. These patterns indicate that both groups demonstrated relatively high grammatical reasoning accuracy overall. However, practitioners exhibited a wider observed accuracy range, including higher maximum scores, whereas student accuracy scores were more tightly clustered.

When considered alongside the summary statistics in Table \ref{comparison_table}, the slightly higher mean and median accuracy observed for practitioners represent a numerical difference rather than a statistically significant group effect. This interpretation is supported by the non-significant Welch's \textit{t}-test result and the small effect size (Cliff’s $\Delta = 0.27$), which together indicate substantial overlap between practitioner and student accuracy distributions.
Taken together, the results in Table \ref{comparison_table} and Figures \ref{His_Prac} and \ref{His_Stu} highlight differences primarily in variability and upper-range accuracy rather than overall grammatical reasoning accuracy levels. This distributional distinction provides a basis for examining how grammatical reasoning accuracy may interact with other constructs, such as personality traits and problem-solving accuracy, in subsequent analyses.\\

\begin{results}
\textbf{Cognitive Capabilities:}
Practitioners and students demonstrated relatively high overall grammatical reasoning accuracy. Practitioners showed slightly higher mean and median accuracy than students; however, this difference was not statistically significant. Differences between groups were primarily evident in score distributions rather than central tendency: practitioners exhibited a wider observed accuracy range, including higher maximum scores, whereas student accuracy scores were more densely distributed.
\end{results}

\subsection{Software Problem-Solving Accuracy}

Accuracy in logic-based reasoning and coding tasks was assessed by participants’ ability to interpret pseudo-code, understand basic data structures and algorithms, and solve analytical problems commonly encountered in software interview settings. Each participant answered six coding-related questions and three logical-reasoning questions. We measured overall problem-solving accuracy by computing the percentage of correct responses for the nine questions.

As shown in Table \ref{coding_logical_comparison}, students achieved higher average accuracy across both coding and logical-reasoning tasks. For coding accuracy, students achieved a mean score of 3.18 (SD = 1.48) out of six, compared to 2.35 (SD = 1.23) for practitioners. A similar pattern was observed for logical reasoning accuracy, where students achieved a mean of 2.05 (SD = 0.93) out of three, compared to 1.44 (SD = 0.99) for practitioners. When combined, total problem-solving accuracy was also higher among students (M = 5.23, SD = 2.02) than practitioners (M = 3.75, SD = 1.94). Median total accuracy further reflects this difference, with students achieving a median accuracy of 55.6\% compared to 33.3\% among practitioners. While both groups reached similar upper bounds of performance, practitioners exhibited a wider observed accuracy range (0–88.9\%) than students (11.1–88.9\%), indicating greater dispersion in practitioner accuracy scores.

\begin{table}[b]
\centering
\footnotesize
\caption{Comparison of coding and logical-reasoning accuracy between practitioners and students.}
\label{coding_logical_comparison}
\begin{tabular}{lcc}
\toprule
\textbf{Measure} & \textbf{Practitioners} & \textbf{Students} \\
\midrule
Coding Accuracy Mean (SD)      & 2.35 (1.23) & 3.18 (1.48) \\
Logical Reasoning Accuracy Mean (SD) & 1.44 (0.99) & 2.05 (0.93) \\
Total Accuracy Mean (SD) /9   & 3.75 (1.94) & 5.23 (2.02) \\
Median (Total Accuracy \%)   & 33.3  & 55.6 \\
Range (Total Accuracy \%)   & 0--88.9 & 11.1--88.9 \\
\midrule
\multicolumn{3}{p{0.9\columnwidth}}{\textbf{Welch’s $t$-tests:} Coding ($p = 0.008$); Logical reasoning ($p = 0.006$); Total accuracy ($p = 0.001$).} \\
\multicolumn{3}{p{0.9\columnwidth}}{\textbf{Effect sizes (Cliff’s $\Delta$):} Coding = $-0.34$; Logical reasoning = $-0.34$; Total = $-0.40$ (small--medium).} \\
\bottomrule
\end{tabular}
\end{table}

Welch’s \textit{t}-tests revealed statistically significant differences between practitioners and students for coding accuracy ($p = 0.008$), logical reasoning accuracy ($p = 0.006$) and total problem-solving accuracy ($p = 0.001$). Effect size analysis using Cliff’s delta indicated small-to-medium effects favoring students across all three measures ((Cliff’s $\Delta$, coding = $-0.34$; logical reasoning = $-0.34$; total = $-0.40$ ). However, substantial overlap in accuracy distributions was observed between the two groups.

\begin{figure}[]
\includegraphics[width=0.95\columnwidth]{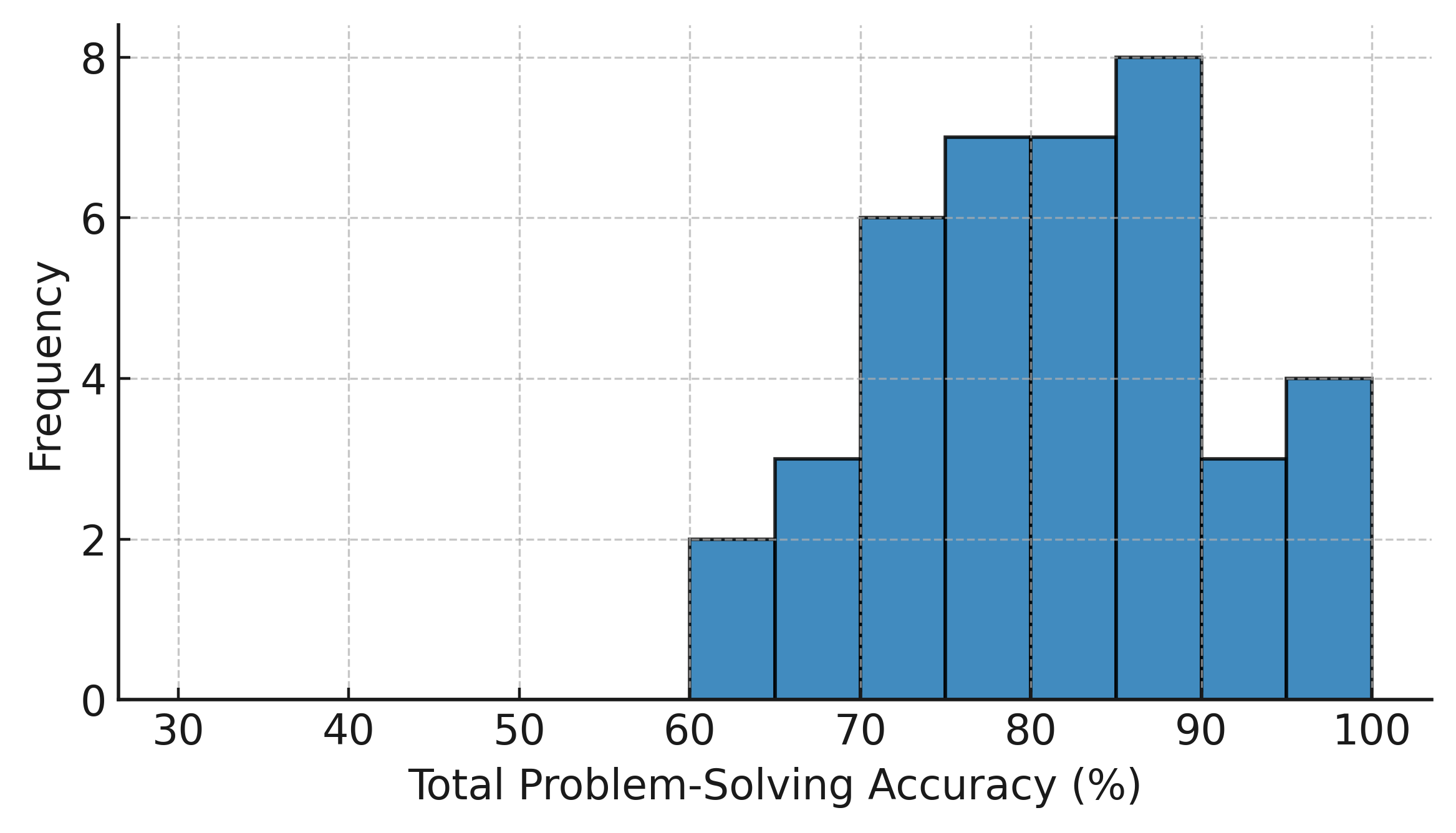}
  \caption{Practitioners' problem solving accuracy.}
  \label{His_Prac_coding}
\end{figure}

\begin{figure}[]
  \includegraphics[width=0.95\columnwidth]{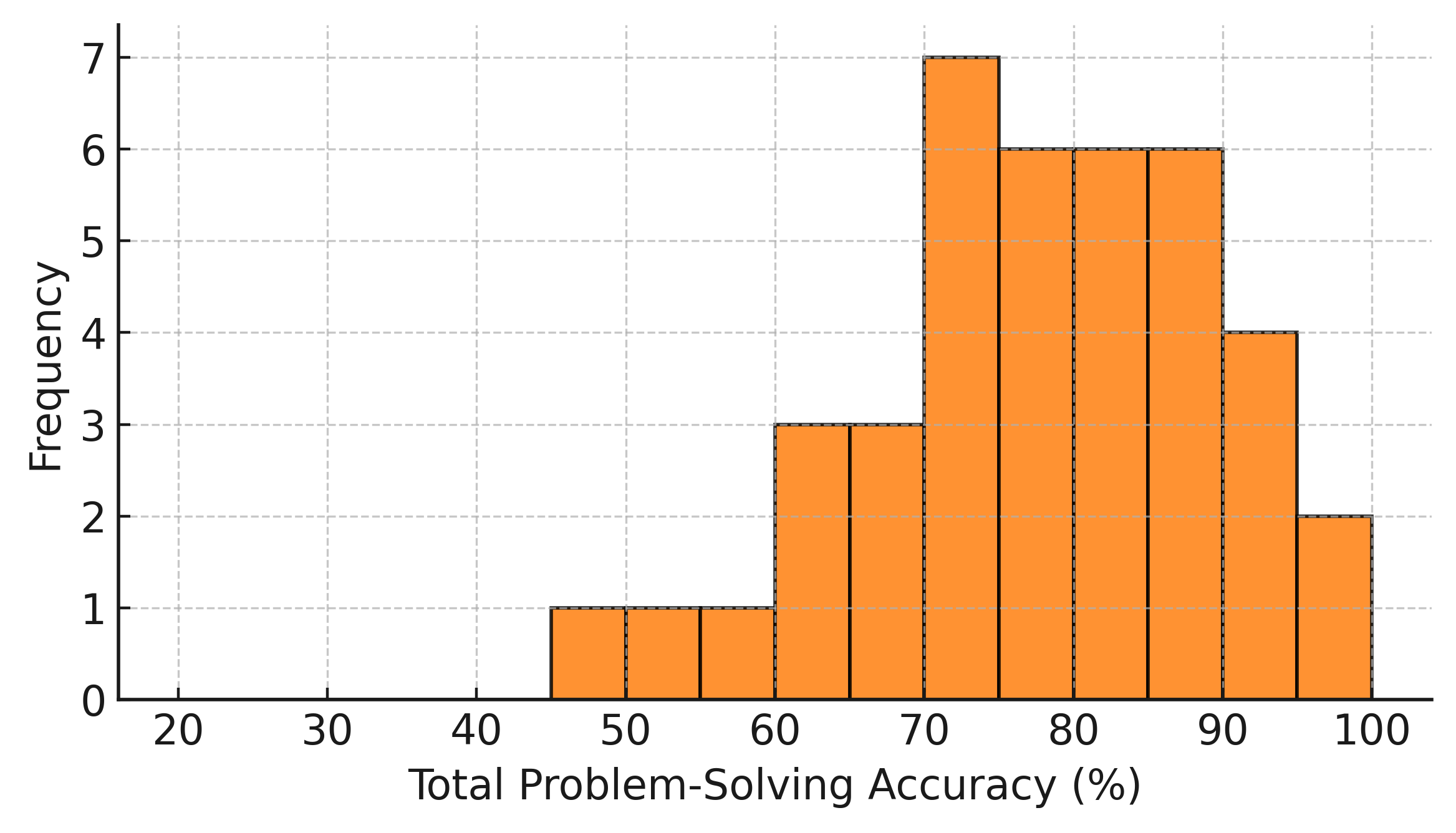}
  \caption{Students' problem solving accuracy.}
  \label{His_Stu_coding}
\end{figure}

The accuracy distributions for practitioners and students are shown in Figure \ref{His_Prac_coding} and Figure \ref{His_Stu_coding}, respectively. Practitioners exhibit a broader and slightly right-skewed distribution, with accuracy scores spread across a wider range and fewer scores concentrated at the higher end. In contrast, students show a more compact distribution centered around the 70–85\% range, reflecting more homogeneous accuracy outcomes across the cohort. These patterns indicate that while both groups demonstrated high problem-solving accuracy, students achieved higher accuracy on the coding and logical reasoning tasks.

Taken together, the results suggest that differences between practitioners and students are characterized more by dispersion and distributional shape than by extreme performance. Students demonstrate higher average accuracy across coding and logical reasoning tasks, while practitioners exhibit a wider spread of outcomes. These findings provide an empirical basis for examining how problem-solving accuracy interacts with other individual characteristics, such as personality traits and grammatical reasoning ability, in subsequent analyses.\\

\begin{results}
\textbf{Problem Solving:}
Students achieved slightly higher average accuracy on the coding, logical reasoning, and combined problem solving tasks compared to practitioners. Median total accuracy was also higher for students, while both groups reached similar upper bounds of accuracy. Practitioners showed a wider accuracy range, whereas student accuracy scores were more densely distributed. Welch's  \textit{t}-tests showed statistically significant differences favoring students across all three measures. Effect sizes were small to medium. % alongside substantial overlap in accuracy distributions between the two groups.
\end{results}

\begin{table}[b]
\centering
\footnotesize
\caption{Pearson correlation between personality traits, grammatical reasoning accuracy, and problem-solving accuracy.}
\label{correlation_table}
\begin{tabular}{p{1.8cm}ccccccc}
\toprule
\textbf{Variable} & \textbf{1} & \textbf{2} & \textbf{3} & \textbf{4} & \textbf{5} & \textbf{6} & \textbf{7} \\
\midrule
1. Extraversion &
-- & 0.21 & 0.09 & -0.12 & 0.18 & 0.05 & 0.11 \\

2. Agreeablen. &
   & -- & 0.34* & -0.19 & 0.22 & 0.08 & 0.13 \\

3. Conscienti. &
   &    & -- & -0.27* & 0.31* & 0.26* & \textbf{0.39**} \\

4. Neuroticism &
   &    &    & -- & -0.29* & -0.18 & -0.23* \\

5. Openness & % to Experience 
   &    &    &    & -- & 0.28* & 0.32* \\

6. Reasoning &
   &    &    &    &    & -- & \textbf{0.41**} \\

7. Problem Solv. &
   &    &    &    &    &    & -- \\
\bottomrule
\end{tabular}

\smallskip
\begin{flushleft}
\footnotesize
\textit{Note.} *$p < .05$, **$p < .01$. Positive correlations indicate direct relationships between traits, reasoning, and problem solving. \textbf{Bold} values indicate correlations significant at $p < .01$.
\end{flushleft}

\end{table}

\subsection{Correlation Analysis}

To examine potential relationships among personality traits, cognitive capability (grammatical reasoning accuracy), and problem-solving accuracy, a Pearson correlation analysis was conducted using the combined dataset of software practitioners and SE students. The analysis included the five personality traits (extraversion, agreeableness, conscientiousness, neuroticism, and openness to experience), grammatical reasoning accuracy, and total problem-solving accuracy derived from the coding and logical reasoning questions. The aim of this analysis was to examine whether individual differences in personality traits or grammatical reasoning accuracy were associated with higher problem-solving accuracy.

Table \ref{correlation_table} summarizes the correlation coefficients and their statistical significance. Overall, the observed associations were small to moderate in magnitude, indicating interpretable but non-deterministic relationships between personality characteristics, cognitive capability (grammatical reasoning accuracy), and problem-solving accuracy.
The strongest association was observed between conscientiousness and problem-solving accuracy (\(r = 0.39, p < .01; \\95\%~\mathrm{CI}~[0.19, 0.56]\)). Individuals reporting higher levels of conscientiousness thus tended to have higher accuracy on coding and logical reasoning questions. Conscientiousness was also positively correlated with grammatical reasoning accuracy (\(r = 0.26, p < .05; 95\%~\mathrm{CI}~[0.04, 0.45]\)), suggesting an association between this trait and grammatical reasoning accuracy. 

Grammatical reasoning accuracy was positively correlated with problem-solving accuracy (\(r = 0.41, p < .01\)), indicating that participants who achieved higher accuracy on the grammatical reasoning task also tended to perform better on the problem-solving tasks. Openness to experience showed positive associations with both grammatical reasoning accuracy (\(r = 0.28, p < .05; 95\%~\mathrm{CI}~[0.06, 0.47]\)) and problem-solving accuracy (\(r = 0.32, p < .05; 95\%~\mathrm{CI}~[0.11, 0.50]\)). This may reflect potential advantages of having characteristics such as curiosity, intellectual flexibility, and open-minded nature when
tackling abstract or unfamiliar programming problems.
In contrast, neuroticism exhibited small negative correlations with grammatical reasoning accuracy (\(r = -0.18, p > .05; 95\%~\mathrm{CI}~[-0.39, 0.04]\)) and problem-solving accuracy (\(r = -0.23, p < .05; 95\%~\mathrm{CI}~[-0.43, -0.01]\)). Although modest in magnitude, this suggests that higher emotional instability may be associated with slightly reduced accuracy, particularly under time-constrained analytical conditions such as our grammatical reasoning tasks. Extraversion and agreeableness showed weak and non-significant associations with grammatical reasoning accuracy and problem-solving accuracy. This suggests limited impact of such interpersonal traits on the accuracy of performing our coding, logical, and grammatical reasoning tasks.

Taken together, these findings suggest that the big five personality traits related to persistence and cognitive flexibility, particularly conscientiousness and openness to experience, may be associated with individual differences in reasoning and problem-solving accuracy. They also suggest that other traits primarily contribute to different personality profiles rather than different performance outcomes. Our results align with prior work indicating that conscientious individuals often exhibit more systematic debugging approaches \cite{kanij2013investigation}, while openness to experience has been linked with creative solution generation \cite{cruz2015forty}. The modest negative associations observed for neuroticism  suggest potential cognitive costs of stress and  uncertainty  on accuracy of performing time-bounded analytical tasks. Overall, our study results suggest that differences in some personality and cognitive capability aspects both contribute to variability in problem-solving accuracy in SE.\\ %This reinforces the potential value of considering individual differences when designing SE task training, recruitment and evaluation frameworks.\\

\begin{results}
\textbf{Correlations:}
The results of our correlation analysis indicate that grammatical reasoning accuracy is positively associated with problem solving accuracy. Among personality traits, conscientiousness and openness to experience show positive associations with both grammatical reasoning accuracy and problem solving accuracy, while neuroticism shows a small negative association with problem solving accuracy. Other traits exhibit weak or non-significant relationships. These findings highlight that variability in problem solving accuracy may be related to individual differences in cognitive capability and personality.
\end{results}

%To facilitate interpretation, the correlation coefficients were visualized using a diverging heatmap (Figure \ref{heatmap}).
%Warmer tones denote positive associations, whereas cooler tones indicate negative relationships. The plot clearly highlights three major trends; 

%\begin{figure}[]
 % \includegraphics[width=0.8\columnwidth]{Correlation_Heatmap.png}
 % \caption{Correlation heatmap among personality traits, cognitive reasoning, and problem solving performance}
 % \label{heatmap}
%\end{figure}

%{\color{red}\textbf{Fix the following, what is it supposed to be?}}

%\par \faLightbulbO\hspace{0.1cm}
%positive clustering between conscientiousness, openness to experience, cognitive capability (reasoning accuracy), and performance

%\par \faLightbulbO\hspace{0.1cm}Mild negative cluster between neuroticism and both cognitive measures, reflecting a negative trend in task accuracy under pressure
%\par \faLightbulbO\hspace{0.1cm}Near-neutral associations for extraversion and agreeableness, confirming limited cognitive influence

\section{Discussion}

Our study examined how cognitive capability and personality traits relate to software problem-solving accuracy, using a comparative analysis of practitioners and students. The two groups did not differ significantly in their personality profiles, with no statistical significance observed across any of the five personality traits. This suggests that observed differences in reasoning and problem-solving accuracy between practitioners and students are unlikely to be explained by systematic group-level differences in personality, even though individual personality traits are associated with accuracy measures within the combined sample. Across the two participant groups, grammatical reasoning accuracy showed a small, non-significant numerical advantage for practitioners, accompanied by greater variability and several high-performing outliers. In contrast, students demonstrated higher and more consistent accuracy on the coding and logical reasoning tasks. Together, these findings suggest that practitioners and students may differ less in baseline cognitive capability and more in how consistently they apply reasoning strategies across different task types. Our study findings have several potential implications for both SE education and professional development. 

\par \faHandORight\hspace{0.1cm} \textbf{Translating conscientiousness into consistent performance}: Conscientiousness emerged as the strongest predictor of both reasoning and problem-solving accuracy at the individual level across the combined sample. Individuals with higher conscientiousness demonstrated greater stability and fewer reasoning errors, consistent with their tendency toward organized, methodical task execution. This association was observed within participants rather than as a difference between practitioners and students, who did not differ significantly in their personality profiles. Embedding reflective checkpoints, such as structured debugging logs or code review planning templates, could help reinforce disciplined work habits. In professional teams, similar strategies may support accountability, predictability, and reliability in performing software tasks.

\par \faHandORight\hspace{0.1cm} \textbf{Bridging the educational and professional reasoning gap}: Our students demonstrated higher and more tightly clustered problem-solving accuracy on the coding and logical reasoning questions, and our practitioners exhibited greater variability, including several high-performing outliers. This pattern suggests differences in consistency rather than overall capability, with students showing more uniform accuracy and practitioners displaying a wider spread of outcomes. Embedding authentic tasks, such as debugging with incomplete specifications or reasoning under uncertainty, may help students engage with a broader range of problem contexts and better prepare for the variability observed among practitioners.

\par \faHandORight\hspace{0.1cm} \textbf{Informing recruitment and role allocation}: Our observed associations between personality traits, grammatical reasoning accuracy, and problem-solving accuracy offer potential guidance for team coordination in software development. Tasks requiring sustained attention and accuracy, such as testing or refactoring, may align well with individuals that have a higher conscientiousness. Tasks involving abstraction or exploratory reasoning, such as architecture design or prototyping, may align with higher openness to experience. These associations can inform task allocation and team coordination by supporting complementary working styles rather than prescribing fixed roles. However, care must be taken not to exclude or constrain individuals based on personality or cognitive profiles. Given the substantial overlap in both personality traits and accuracy measures, training and team support remain critical for enabling individuals with different personality traits and cognitive capabilities to perform effectively across different SE tasks.

\par \faHandORight\hspace{0.1cm} \textbf{Implications for researchers}: This study serves as a preliminary step toward understanding how personality traits and cognitive capability jointly relate to software problem-solving accuracy. Future research could expand these findings through larger, longitudinal, and cross-cultural samples, as well as by incorporating additional measures such as physiological or behavioral indicators of cognitive load. In addition, experimental studies could explicitly examine causal mechanisms, for example by investigating how training interventions or team composition influence the relationships between conscientiousness, openness to experience, grammatical reasoning accuracy, and problem-solving accuracy. As an initial effort to quantitatively link these constructs using both practitioner and student populations, this study provides a foundation for a growing line of inquiry into the human dynamics shaping analytical competence in SE.

\section{Limitations and Threats to Validity}
%We acknowledge the following limitations of our study.

\textit{Construct validity:} Personality was measured using the IPIP-NEO-50, a widely used and validated self-report instrument. As with all self-reported measures, responses may be affected by self-perception bias. Although the FFM provides a robust framework, it captures broad traits and may not fully reflect task-specific behaviors relevant to SE. Cognitive capability was assessed using a grammatical reasoning test, which emphasizes verbal reasoning and does not directly capture other forms of reasoning used in software development (e.g., algorithmic or debugging reasoning). To address this, applied problem solving was assessed separately using coding and logical-reasoning tasks, as described in the Section \ref{coding_logical_questions}.

\textit{Internal validity:} The study employed a correlational design; therefore, causal relationships between personality, cognitive capability, and problem-solving accuracy cannot be inferred. Although statistically significant associations were observed, unmeasured factors such as prior exposure to similar tasks, or individual test-taking strategies may have influenced outcomes. Given the language-dependent nature of the grammatical reasoning task, variation in English proficiency among non-native participants, although generally reported as native-like or advanced, may have affected their reasoning accuracy. English proficiency was self-reported and not controlled for in the analysis. 

\textit{External validity:} Our study population sample comprised 40 practitioners and 40 students recruited from multiple countries, providing a degree of geographic diversity. However, participants were primarily drawn from technology-oriented and English-speaking contexts, which may limit the generalizability of the findings to other populations. % operating in different linguistic, educational, or cultural environments.
In addition, the practitioner group included participants from a range of roles and application domains. Although this reflects the heterogeneity of contemporary software practice, differences in task specialization, professional responsibilities, or depth of experience were not explicitly modeled. As a result, the findings should be generalized with caution. %, particularly to contexts where software development practices, educational pathways, or communication norms differ substantially from those represented in the sample.

\textit{Statistical conclusion validity: }The study employed descriptive statistics, group comparisons, and Pearson correlations to examine relationships among variables. Although statistically significant associations were identified, the analyses were limited to bivariate relationships, with no multivariate modeling or control for potential confounders. Accordingly, the findings should be interpreted as exploratory associations rather than evidence of complex or conditional relationships among personality traits, cognitive capability, and problem-solving accuracy.

These limitations do not undermine the value of our findings, rather highlight the need for more nuanced, multi-method research. % Future work could replicate our design with larger samples, incorporate other behavioral or physiological measures of reasoning, and explore the longitudinal evolution of cognitive- personality dynamics across educational and professional transitions in SE.

\section{Conclusion}

Our study indicates that software problem-solving accuracy is associated with both individual personality characteristics and cognitive capability. Across a sample of 40 software practitioners and 40 students, conscientiousness and openness to experience were positively associated with grammatical reasoning accuracy and problem solving accuracy, while neuroticism showed a small negative association. These findings represent an initial step toward understanding how personality, and cognitive capability, jointly relate to software problem solving accuracy. Future work could extend these findings through larger, longitudinal, and cross-cultural studies, incorporate behavioral or physiological indicators of cognitive load, and examine how training interventions or team composition may moderate the relationships between personality traits, reasoning accuracy, and problem solving outcomes.

\begin{acks}
Hidellaarachchi and Grundy were supported by ARC Laureate \\{Fellowship} FL190100035.  
\end{acks}

\bibliographystyle{ACM-Reference-Format}
\bibliography{reference}

\appendix

\end{document}